\documentclass{emulateapj}
\usepackage{psfig}
\usepackage{graphicx}
\usepackage{amsmath}
\DeclareGraphicsExtensions{.png,.jpg,.pdf}

\newcommand{\beq}{\begin{equation}} 
\newcommand{\seq}{\end{equation}} 
\newcommand{\pdif}[2]{\ensuremath{ \frac{\partial #1}{\partial #2}}} 
\newcommand{\f}[2]{\frac{#1}{#2}} 

\begin{document}
\title{CLOUD FORMATION AND ACCELERATION IN A RADIATIVE ENVIRONMENT}
\author{Daniel Proga\altaffilmark{1} and Tim Waters\altaffilmark{1}}
\altaffiltext{1}{Department of Physics \& Astronomy, University of Nevada, 
Las Vegas}

\begin{abstract}
In a radiatively heated and cooled medium, the thermal instability is a
plausible mechanism for forming clouds, while the radiation force provides
a natural acceleration, especially when ions recombine and opacity increases. 
Here we extend Field's theory to self-consistently account for a radiation 
force resulting from bound-free and bound-bound transitions in the optically 
thin limit.  We present physical arguments for clouds to be significantly 
accelerated by a radiation force due to lines during a nonlinear phase 
of the instability. To qualitatively illustrate our main points, we perform 
both one and two-dimensional (1-D/2-D) hydrodynamical simulations that allow 
us to study the nonlinear outcome of the evolution of thermally unstable gas 
subjected to this radiation force. Our 1-D simulations demonstrate that 
the thermal instability can produce long-lived clouds that reach 
a thermal equilibrium between radiative processes and thermal conduction,
while the radiation force can indeed accelerate the clouds to supersonic velocities. 
However, our 2-D simulations reveal that a single cloud with a simple 
morphology cannot be maintained due to destructive processes, triggered by 
the Rayleigh-Taylor instability and followed by the Kelvin-Helmholtz instability.
Nevertheless, the resulting cold gas structures are still significantly
accelerated before they are ultimately dispersed.
\end{abstract}


\section{Introduction}
The thermal stability of astrophysical gases has been extensively investigated,
both analytically and numerically, in many different physical contexts, 
e.g., in star formation regions, in gas accreted by forming galaxies, 
and in interstellar and intergalactic media. The physical basis for gas 
in thermal equilibrium to be subject to thermal instability (TI) was 
introduced in a classic paper by Field (1965), who developed the linear theory 
accounting for thermal conduction. Later Balbus (1986) generalized Field's
theory by relaxing the equilibrium assumption and permitting the background 
flow to be time-dependent.

One specific instance where gas can be prone to TI is when both cooling
and heating are due to radiative processes. This situation occurs for example 
in active galactic nuclei (AGN) where, for a temperature of about 
$10^5-10^6$~K, the cooling is dominated by free-free emission while 
the energy gain is dominated by X-ray absorption and Compton scattering of energetic photons 
produced by an accreting black hole (e.g., Krolik et al. 1981, Lepp et al. 1985).
In this situation, the radiation field is almost in the free-streaming limit
and consequently the gas will receive a non-zero net push
in a direction away from the source of radiation.  Our objective here 
is to demonstrate that this push can be substantial due to
a large increase in the bound-free and bound-bound opacities of the cold
gas during a nonlinear phase of the TI.

In this paper, we present results from our
detailed study on the nonlinear dynamics of thermally unstable gas that 
is accelerated by the same photons that establish its thermal environment.  
In \S{2}, we describe the theoretical underpinnings on which this paper 
is premised. In \S{3}, we outline the equations solved and the explicit 
form of both the energy and momentum source terms.  
In \S{4}, we present our methods and results from 1-D and 2-D 
simulations.  We finish in \S{5} where we discuss our findings.

\begin{figure}
\includegraphics[width=0.5\textwidth]{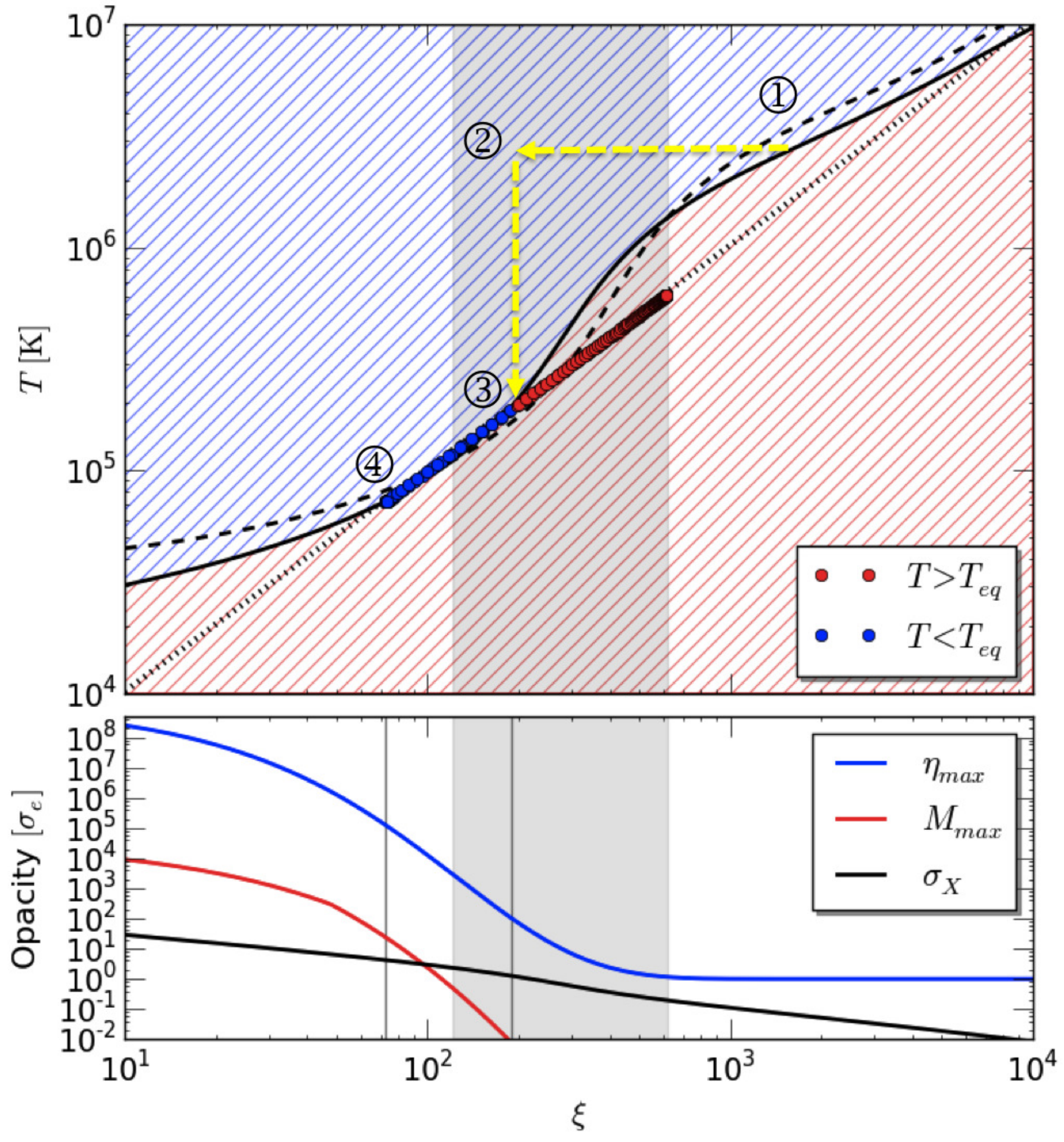} 
\caption{Temperature and opacity dependence on the photoionization parameter expected in an AGN environment.  
\textbf{Top panel}: The solid line is the S-curve found by solving $\mathcal L(T,\xi)=0$.  
The region above (below) this line is patterned light blue (red) to denote cooling (heating), 
as gas in this region is above (below) the equilibrium temperature. The dashed curve (defined by 
$[\partial \mathcal{L}/T)/\partial T]_p =0$) marks the isobaric instability criterion.  Thermally stable 
gas must lie below this curve; gas anywhere in the grey region cannot settle on the S-curve in this 
region without being unstable.  The dotted line shows a constant pressure slope.  Stable gas at 
location (1) is envisioned to be suddenly subjected to a reduced flux, placing it at location (2), 
where it is unstable.  This gas will rapidly cool (nearly isochorically) until it reaches a new
thermal equilibrium which is now unstable [marked as location (3) on the equilibrium curve]. 
We begin our simulations at this new equilibrium to follow the growth of an  
isobaric perturbation.  As one can anticipate, the perturbation grows maintaining pressure equilibrium
even during the non-linear phase and the points representing gas 
move along the dotted line; in particular those representing the cold gas move toward 
yet another thermal equilibrium location (4) which is now stable.  
All points in the computational domain of 1-D run RFLDX (radiation force due to X-rays and lines) 
are over-plotted as blue $(T<T_{eq})$ and red $(T>T_{eq})$ dots to indicate the final state of the gas. 
\textbf{Bottom panel}: Gas opacity in the units of the Thomson
opacity as a function of $\xi$.
The solid red line represents
bound-bound opacity, $M_{\rm max}$.
This opacity 
can become orders of magnitude larger than bound-free opacity $\sigma_X$ (shown as the solid black line). 
The solid blue line represents the opacity of the most opaque line.
To gauge the increase in 
opacity for the cloud formed in the top panel [location (4)], 
the two vertical lines mark the initial conditions ($\xi_3=190$) 
of the gas and the eventual location of the cloud core ($\xi_4 \approx 73$).  }
\label{fig:T-xi_plot}
\end{figure}

\section{Theoretical Model/Expectations}
\label{microphysics} 
To facilitate a simple comparison between our simulations and Field's theory 
(i.e. the linear growth rates), we include thermal conduction, assume 
the gas to be initially stationary, homogenous, and non-magnetized, 
and we do not include gravity. 
We model the formation and evolution of optically thin clouds whose thermal 
equilibrium state is controlled by impinging radiation by adopting a realistic prescription for heating and cooling 
appropriate for gas in an AGN environment.  Specifically, we use a net cooling 
function, $\mathcal L$, that includes the following radiative processes: 
(1) Compton and inverse-Compton scattering,
(2) photoionization and recombination, 
(3) Bremstrahlung, and
(4) line-emission. In an optically thin case,  $\mathcal L$ depends on the gas temperature, $T$,
and photoionization parameter, $\xi = 4\pi \mathcal{F}_{X}/n$, where $\mathcal{F}_{X}$ is the X-ray flux
and $n$ is the gas number density.\footnote{The units for $\xi$ are $erg~cm~s^{-1}$ and we do not
cite them throughout the remaining part of this paper.}

The following theoretical picture forms the basis of our expectations 
and motivates our simulation setup.  
For a given $\xi$, gas in equilibrium at temperature $T_{eq}$ will have heating 
balancing cooling [i.e., $\mathcal{L}(T_{eq},\xi)=0]$. 
Therefore, it will occupy one point on 
the radiative equilibrium curve (the solid line which is hereafter denoted 
the S-curve) shown in the top panel of Fig.~1.
Suppose that initially this point 
is at a stable location on the S-curve such as at position 1 in Fig.~1, 
but some physical event transpires that results in a reduction of $\mathcal{F}_X$ 
so that the gas finds itself out of equilibrium 
at location 2 with, for example, $\xi_2 = 190$.  It will quickly cool 
to reside on the S-curve at location 3 (with $\xi_3 = \xi_2$).  
However,
at this position, gas is thermally unstable to perturbations with constant pressure $p$,
as it violates Field's criterion for stability, $[\partial \mathcal L/\partial T]_p >0$.  
More generally, only the region below the dashed line is 
thermally stable according to Balbus' criterion 
$[\partial (\mathcal L/T)/\partial T]_p >0$, 
with the consequence that gas occupying the shaded grey region 
in Fig.~1 will evolve to points on the S-curve outside of the grey region.  
Continuing with our example, the slightest density perturbation present 
in the gas at location 3 will grow exponentially at the linear theory rate 
and at nearly constant pressure until the perturbation becomes nonlinear.  
Rapid nonlinear growth (still at nearly constant pressure) will then commence 
(e.g., Burkert \& Lin 2000) and result in the formation of a cloud with 
a core density approximately determined by the intersection of the dotted line 
and the S-curve (position 4), which is where the gas in the core is stable.

The bottom panel of Fig.~1 illustrates how various gas opacities depend on $\xi$.
Solid black and red lines show the bound-free opacity, $\sigma_X$,
and the total line opacity, $M_{\rm max}$, respectively. The solid blue line
represents the opacity of the most opaque line. One can see 
that for $\xi_3$, the bound-free and bound-bound opacity 
is negligible.  
(The vertical line in the grey region marks the location $\xi_3 = 190$.)
However, once the thermally unstable gas forms a dense cloud, 
$\sigma_X$ and $M_{\rm max}$
are significantly increased, as indicated by the left vertical line at $\xi_4 \approx 73$.
Therefore the cloud can be  accelerated by the same source that heats it.

\section{Governing Equations}
\label{equations}

To confirm and quantify our expectations, we numerically solve the equations of
hydrodynamics:
\begin{equation}
\pdif{\rho}{t} + \mathbf{\nabla} \cdot \left(\rho \mathbf{v} \right) = 0 \label{eq:mass} ,
\end{equation}

\begin{equation}
\pdif{\left(\rho\mathbf{v}\right)}{t} + \mathbf{\nabla} \cdot \left( \rho\mathbf{v} 
\mathbf{v} + p\,{\sf I}\right) = \mathbf{f}_{rad},\label{eq:mom}
\end{equation}

\begin{equation}
\pdif{E}{t} + \mathbf{\nabla} \cdot [(E + p)\mathbf{v}] =
-\rho {\mathcal L} + \kappa_0\nabla^2T.
\label{eq:energy}
\end{equation}
In the above, $\rho$ is the gas density, $\mathbf{v}$ is the fluid velocity, $p$ is the gas pressure, 
$\sf I$ is the unit tensor, $\mathbf{f}_{rad}$ is 
the radiation force (see below), and $E = e + \rho v^2/2$ is the total energy density, with 
$e$ the internal energy density.  
This system is closed with an ideal gas equation of state, $e=p/(\gamma-1)$, using an adiabatic index 
$\gamma = 5/3$.  We estimate the conduction coefficient $\kappa_0$ using 
the value for a fully ionized plasma (Spitzer 1962) evaluated at 
the equilibrium temperature and density of position 3 in Fig.~1 
(hereafter a suffix `eq' denotes evaluation of a quantity at $[\xi_3, T_{eq}]$).

We denote the functional dependence of the net cooling function as
$ {\mathcal L} =   \Lambda - \Gamma$, with 
\begin{align}
 \Lambda &=   \f{n}{\mu m_p} (L_{\rm ff} + L_{\rm bb}) \, \, {\rm [erg\,g^{-1}\, s^{-1} ]}, \label{eq:Lambda} \text{ and}\\
 \Gamma &=   \f{n}{\mu m_p} (G_{C} + G_X) \, \, {\rm [erg\,g^{-1}\, s^{-1} ]}, \label{eq:Gamma}
\end{align}
where $\mu$ is the mean molecular weight (set to 1.0 in this work) such that 
$n = \mu m_p \rho$ (with $m_p$ the proton mass), 
$L_{\rm ff}$ and $L_{\rm bb}$ are the cooling rates due to free-free and bound-bound 
transitions, and $G_C$ and $G_X$ are 
the heating rates due to Compton and X-ray heating, respectively
(all four rates are  in units of $\rm erg\,\, cm^3~s^{-1}$).  For $L_{\rm ff}$ and $G_C$, we use 
the well-known analytic formulas based on atomic physics; see the appendix.
For $L_{\rm bb}$ and $G_X$, meanwhile, we use the analytical fits given by Blondin (1994), 
who found good agreement (to within 25\%) between his approximate rates and 
those resulting from detailed photoionization calculations.   Blondin's formulae, also provided in the appendix, 
assume an optically thin gas of cosmic abundances illuminated by a $T_X = 10\,\rm{keV}/k_B$ bremsstrahlung 
spectrum (with $k_B$ Boltzmann's constant).

To evaluate $\mathbf{f}_{rad}$, we consider the source of radiation to be 
located at $x=-\infty$, giving
\beq
\mathbf{f}_{rad} = \f{\rho\sigma_{tot} \mathcal{F}_{tot}}{c} \hat{x} \, \, {\rm [g\,cm^{-2}\, s^{-2} ]}, \seq
where $\mathcal{F}_{tot}$ is the total continuum flux, $c$ is the speed 
of light, and $\sigma_{tot}$ is an effective total opacity, individual
contributions to which can be self-consistently 
estimated as we describe below.

The X-ray flux is set by the photoionization parameter $\xi$. Therefore,
we can specify the product $\sigma_{tot}\mathcal{F}_{tot}$ by assuming that
there is nonzero continuum opacity only for X-rays and nonzero line opacity
for UV photons. With this assumption we
parametrize the UV flux $\mathcal{F}_{\rm{UV}}$ in terms of the X-ray flux 
using $f_{\rm{UV}} \equiv \mathcal{F}_{\rm{UV}}/\mathcal{F}_{X}$ and use the following expression
to estimate the opacity and flux product:
 \beq \sigma_{tot}\mathcal{F}_{tot} = \sigma_e\Big[(1+f_{\rm{UV}}) + \sigma_X  + f_{\rm{UV}}M_{\rm max}\Big] \mathcal{F}_{X}. \label{eq:sigma_tot}\seq
Here, $\sigma_e$ is the mass scattering coefficient of the free electrons.
The term $(1+f_{\rm{UV}})$ in equation 
(\ref{eq:sigma_tot}) is due to electron scattering, while the term 
$\sigma_X$ is an effective X-ray opacity in units of $\sigma_e$.  
We compute $\sigma_{X}$ as $(4\pi G_{X,h}/\xi)/\sigma_e$, where $G_{X,h}$ is the heating part 
of $G_X$ (see the appendix).  The last term in 
equation (\ref{eq:sigma_tot}), $M_{\rm max}$, 
is the maximum force multiplier characterizing the line contribution 
to the total opacity. We evaluate $M_{\rm max}$ following
Stevens \& Kallman (1990, SK90 hereafter) 
with a modification, as described below. 
Note that in AGN, $f_{\rm{UV}}>1$. 

The maximum force multiplier parametrizes the increase in the scattering 
coefficient due to spectral lines when all of the lines are optically thin.  
In this optically thin limit, the multiplier is just a sum of opacity
contributions from all the lines and 
depends only on 
the gas composition, ionization, excitation and oscillator strengths
[e.g., see eqs. 10 and 11 in Castor, Abbott, \& Klein, (1975); CAK hereafter].
Following  Owocki, Castor, \& Rybicki (1988), who used a modified CAK 
method, one can evaluate
this maximum multiplier as
\beq M_{\rm max} = k_{\rm{CAK}}(1-\alpha)\eta_{\rm max}^\alpha, \seq
where $k_{\rm CAK}$ is proportional to the total number of lines,
$\alpha$ is the ratio of strong to weak lines, and finally 
$\eta_{\rm max} = \kappa_{L,\rm max}/\sigma_e$ 
is a dimensionless measure of opacity of the most opaque line
(with $\kappa_{L,\rm max}$ being the line opacity coefficient of 
the thickest line).

SK90 carried out detailed photoionization calculations for a radiative environment appropriate for 
X-ray sources and parametrized their results 
in terms of the above expression for $M_{\rm max}$ by allowing $k_{\rm{CAK}}$ to be 
$T$-dependent and $\eta_{\rm max}$ to be $\xi$-dependent. 
Instead of the fit for $k_{\rm{CAK}}(T)$ from SK90, we use equation (17) of Proga (2007) 
due to Kallman (2006, private communication), as this may better represent 
the increase in the number of lines with decreasing temperature in AGN.  
This expression and that for $\eta_{\rm max}(\xi)$, which is equation (19) of SK90, are provided 
in the appendix.  
Both of these fits were generated assuming $\alpha = 0.6$.
In the bottom panel of Fig.~1, we plot $\sigma_X$ along with $M_{\rm max}$.  Notice that $M_{\rm max}$ 
can be roughly a few thousand for gas ionized by a weak radiation field, 
whereas it decreases asymptotically to zero for highly ionized gas as the radiation field becomes stronger.  

\section{Methods and Results}
We solve equations (\ref{eq:mass})-(\ref{eq:energy}) in 1-D and 2-D using 
the CTU integrator, Roe flux, and explicit conduction module of the MHD code 
\textsc{Athena} (Stone et al. 2008). 
We modify the original version of the code by adding the momentum and heating 
and cooling source terms in the same way that \textsc{Athena}'s built-in
static gravitational potential source term is implemented to achieve 2nd 
order accuracy in both space and time.  
We use a less accurate method 
for integrating the conduction term in time in our 2-D simulations than in our 1-D simulations. 
Specifically, we use \textsc{Athena}'s super time-stepping scheme (STS; see O'Sullivan \& Downes 2006),
although we note that a 2nd order accurate in time STS algorithm does exist (Meyer et al. 2012) and 
we are testing it for future use.

\subsection{Initial and Boundary Conditions}
Given the atomic physics behind our S-curve and the opacities in Fig.~1, the following free parameters 
govern our problem: the wavenumber and density amplitude of the TI perturbation, $k$ and $\delta \rho$, 
as well as the ratio of the initial sound crossing and thermal times $t_{th}/t_{sc}$, which together 
determine the number of clouds 
and their formation time; the initial photoionization parameter $\xi_3$, which controls the intensity of 
the radiation field and the equilibrium temperature $T_{eq}$; the equilibrium pressure, namely the product 
$n_{eq}T_{eq}$, which sets the physical units of the cloud and its environment; and finally $f_{UV}$, 
which parametrizes the shape of the spectral energy distribution in a simple way.  

Our initial conditions are full wavelength profiles of the TI condensation 
mode, found from equations (11)-(14) in Field (1965), applied to density, velocity, and pressure. 
We adopt $n_{eq} T_{eq} = 10^{13} ~\rm{K~cm^{-3}}$ in accordance with AGN observations and their modeling 
(e.g., Davidson \& Netzer 1979; Krolik et al. 1981).  
The length scale of the perturbation is fixed by the adiabatic sound speed at position 3 in Fig.~1, 
$c_{eq}$, and a choice for the initial sound crossing time, $t_{sc}$.  
We chose $t_{sc}$ equal to the initial thermal time, 
$t_{th} = (e_{eq}/\rho_{eq})/\Lambda_{eq}$, 
which results in near maximum linear growth rates 
for the condensation mode.  These rates are obtained by solving 
the dispersion relation in Field (1965; eq. 15).
For both 1-D and 2-D simulations, we use periodic boundary conditions and 
set the domain size in the x-direction, $\Delta x$, equal to the perturbation wavelength 
$\lambda_x = c_{eq} t_{sc}$. 
The amplitude of the density perturbation is $\delta \rho = 5.0\times10^{-5}\rho_{eq}$.

With this setup, the unstable region of Fig.~1 is parametrized entirely by $\xi_3$, and varying 
this parameter leads to the formation of clouds with substantially different properties.   
A realistic value for AGN is likely $\xi_3 = 500$, as this results in a pressure photoionization parameter 
$\Xi \equiv (\mathcal{F}_X/c)/(n_{eq}k_BT_{eq}) \approx 9.0$, and the AGN environment is expected to be 
hospitable to clouds for $\Xi \lesssim 10$ (e.g., Krolik et al. 1981).   For $\xi_3 = 500$, the linear 
growth rate of the TI is comparatively small, taking more than 400 days for the density of the cold gas 
to double, and estimates based on Fig.~1 indicate that a 1-D (or 2-D planar) cloud will form 
at $\xi_c \approx 20$ with a width $l_c\sim0.5\,\rm AU$, a temperature $T_c\sim4\times10^4\,\rm K$, 
a number density $n_c \sim 3\times 10^8\,\rm cm^{-3}$, and a density contrast of $\chi \sim30$.  
The radiation force due to lines can be very powerful, with $M_{\rm max}$ at least $10^3$.  

This more  realistic cloud is, however, very optically thick for many strong UV lines.
(Moreover, as we discuss in \S{5}, it is challenging to accurately resolve in multi-dimensions.)
Our present simulations are designed to explore the optically thin regime, 
as this regime allows us to focus on the purely hydrodynamical effects of cloud formation and 
acceleration without the further complications involved when solving the equations of radiation 
hydrodynamics (RHD; cf. Proga et al. 2014).   
This restricts $\xi_3$ to a very narrow range of values corresponding to larger, less realistic 
values of $\Xi$, as there is obviously an upper limit on the density of clouds whose acceleration 
we can accurately model without RHD.

To estimate this upper limit on the density, we assume the cloud forms with both a constant density $\rho_c$ 
and width $l_c$ (which will be seen to be very nearly the case here), so that we can write 
$\eta_{\rm max} = \tau_{L,\rm max}/\tau_{s}$, where $\tau_{L,\rm max} = \kappa_{L,\rm max} \rho_c l_c$ 
is optical depth of the thickest line and $\tau_{s} = \sigma_e \rho_c l_c$ is the electron scattering 
optical depth of the cloud.  Demanding that the cloud to be optically thin to all bound-bound transitions 
(i.e. $\tau_{L,\rm max} < 1$) requires $\eta_{\rm max} \tau_{s} < 1$, or
\beq \rho_c < \left( \sigma_e l_c \eta_{\rm max} \right)^{-1}.\label{eq:dmax} \seq 
We chose the value $\xi_3 = 190$ since it produces a cloud with the highest density contrast that 
satisfies this inequality at all times in 1-D.  (In \S{4.4}, we verify our optically thin assumption 
in both 1-D and 2-D using a more accurate estimate.)  
For $\xi_3 = 190$, $M_{\rm max}$ does not exceed 40.  To explore the effects of the stronger line force, 
we set $f_{UV} = 10$, which is guided by observational results from 
Zheng et al. (1997) and Laor et al. (1997).  

\subsection{Simulations} 
We have performed over 30 simulations exploring a variety of parameters and 
numerical setups.
Here we report in some detail on a set of four simulations that differ only by the applied radiation force:  
RF (electron scattering only), 
RFX (electron scattering plus X-ray absorption), 
RFLD (electron scattering plus line-driving), and 
RFLDX (electron scattering, line-driving, and X-ray absorption).  
We will especially focus on run RFLDX, as this case is most representative of the physical conditions 
in AGN.  
Compared to a more realistic AGN cloud described above, for our adopted value of $\xi_3 = 190$ 
the cloud growth is much faster, taking only 2.4 days to double in density, while 
$l_c \approx 6.5\times10^{10}\,\rm cm$ ($0.004\,\rm AU$), $T_c \approx 7.0\times10^4\,\rm K$, 
$n_c \approx 1.4\times 10^8\,\rm cm^{-3}$, $\chi \approx 8.0$, and $\Xi \approx 19$.
The physical units corresponding to our numerical results are listed in Table~1. 

\begin{table}[htp]
\centering
\caption{Physical units\\ (corresponding to $\xi=190$ \& $t_{th}/t_{sc} = 1$)}
\begin{tabular}{ll}
\hline
Quantity              & Value (cgs units) \\
\hline
$\rho_{eq}$ & $ 8.65\times10^{-17}\,\rm{g~cm^{-3}}$ \\
$n_{eq}$ & $ 5.17\times10^{7}\,\rm{cm^{-3}}$ \\
$T_{eq}$ & $1.93\times10^5~\rm{K}$ \\
$c_{eq}$&  $5.16 \times 10^6\rm{~cm~s^{-1}}$ \\
$t_{sc}$ & $6.03\times10^4\rm{~s}$ \\
$\lambda_x$ & $3.11\times 10^{11}\rm{~cm}$ \\ 
$\mathcal{F}_X$ &  $7.82\times10^8 \rm{~erg~s^{-1}~cm^{-2}}$ \\
$\kappa_0$ & $1.58 \times 10^7 \rm{~erg~s^{-1}~K^{-1}~cm^{-1}}$ \\
$\Lambda_{eq}$ & $3.97 \times 10^{8} \rm{~erg~g^{-1}~s^{-1}}$ \\
\hline
\end{tabular}
\label{Table:parameters}
\end{table}

\begin{figure*}
\centering
\includegraphics[width=0.75\textwidth]{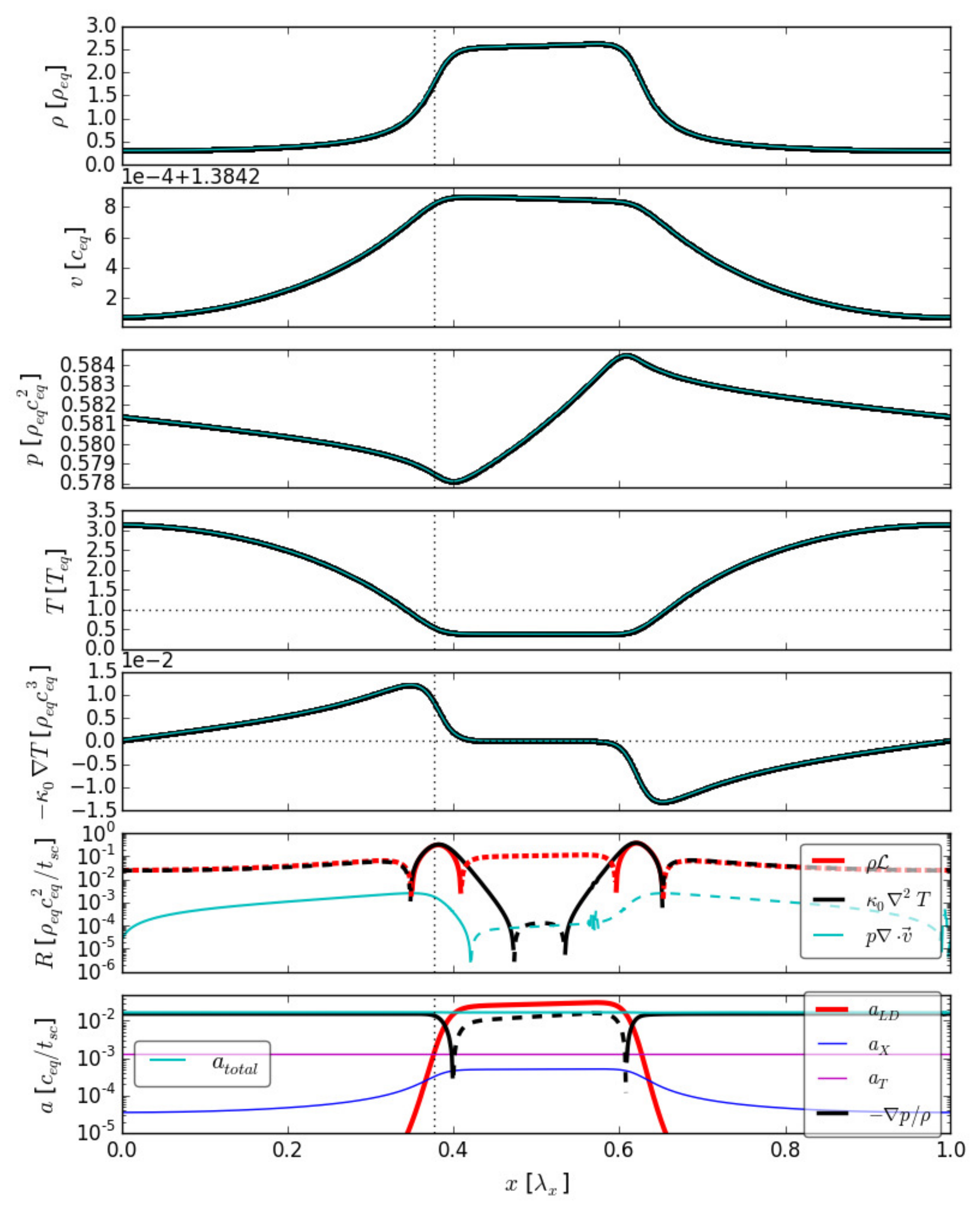}
\caption{Profiles of run RFLDX in 1-D at time $120\,t_{sc}$.  The resolution is $N_x = 1024$ zones.  
Numbers above panels are offsets, e.g., the velocity ranges from about 
$1.3843\,c_{eq}$ to $1.3851\,c_{eq}$.  
The dotted vertical line indicates the position of the maximum density gradient.  
The 2nd from bottom panel 
compares the heating and cooling rates $R$ in the energy equation, while the bottom panel compares 
the various accelerations.  Solid (dashed) portions of lines in these panels indicate positive (negative) 
values, e.g. conduction transfers heat into the interfaces at the expense of the medium, 
while the specific 
pressure force points to the left (opposite the cloud motion) in the cloud core and 
to the right elsewhere.} 
\label{fig:panels}
\end{figure*}

\subsection{Results of 1-D Simulations}
Despite the different radiation forces, the clouds in all four runs are formed
at the same time and with the same density and temperature contrasts, and the gas 
therefore traces the same `tracks' on the $T-\xi$ plot in Fig.~1.  
The over-plotted red and blue dots in Fig.~1 show the tracks for RFLDX 
at $t=120\, t_{sc}$, which represents the time where the flow
reached a thermal steady state (see below for more details).
Note that the red tracks do not reach the radiative equilibrium curve 
(i.e. $\mathcal{L}=0$), but rather an equilibrium curve given by 
$\rho \mathcal{L} = \kappa_0 \nabla^2 T$.  

Also note that there are tracks occupying an unstable (according to Balbus' criterion) region in Fig.~1, namely, the tracks within 
the grey region that are above the dashed line.  Given that the gas in the cloud core occupies location 4 
and is in pressure equilibrium with the medium, it must be the case that some portion of the 
gas occupies this unstable region in order for the density and temperature to be continuous everywhere.    
These `unstable' tracks correspond to the gas in the conductive interfaces of the cloud.  
In Fig.~2, we plot profiles of the solution overplotted in Fig.~1, and the width of the interfaces can 
be judged from the density panel.  
The local Field length in the interfaces is close to the initial equilibrium value, 
$\lambda_F = 2\pi\sqrt{\kappa_{eq}T_{eq}/(\rho_{eq}\Lambda_{eq})}\,\lambda_x \approx 0.19\, \lambda_x$, 
while the interface width is two or three times smaller than this.  
Gas is permitted to be thermally unstable at regions smaller than the Field length when the stabilizing 
influence of conduction 
(measured by the heat flux that is shown in the fifth panel of Fig. 2) is large enough.  
 
We find that in all of these runs the gas arrives at a state of thermal 
equilibrium by $t=90\, t_{sc}$.
The second from bottom panel in Fig.~2 shows how 
this equilibrium state is possible.  Both the net cooling function (red curve) 
and the conduction term (black curve) are positive at the interfaces of 
the cloud and negative in the hot medium.  These terms are of opposite sign 
in equation (\ref{eq:energy}) and therefore balance each other.  
The compression term (cyan) is negligible at this time, but it was 
the dominant term when the cloud was forming.   

The radiation force prevents the gas from reaching a mechanical equilibrium state.   
Rather, in each case the cloud core undergoes dynamical changes (i.e. the pressure 
and velocity profiles adjust) 
to permit nearly uniform acceleration.  To show this, we plot the net flow acceleration 
in the bottom panel of Fig.~2 (cyan line), 
which is the sum of the other curves displayed.  Line driving operates almost uniformly 
throughout the cloud core.  As can be seen by either the acceleration or the pressure panel, 
the response of the gas pressure is to exert a nearly constant drag force on the cloud to 
compensate for the driving force, while the medium is pushed along since it has nowhere else 
it can go in 1-D. The adjustment of  the forces is obtained shortly after the cloud is formed, 
with the acceleration profiles resembling those shown in the bottom panel at around 
$t = 55\, t_{sc}$ for runs RFLD and RFLDX. Some profiles (especially velocity) continue 
to undergo changes until $t=120\, t_{sc}$; the shapes 
shown in Fig.~2 are maintained as the cloud continues to accelerate to high Mach numbers.

The results from our 1-D simulations confirm our basic picture for cloud formation 
and acceleration.  As a next step, we perform 2-D simulations where 
destructive processes may change our results.  Our primary concern is 
the Rayleigh-Taylor (RT) instability, and subsequently the Kelvin-Helmholtz 
(KH) instability.  Fig.~2 shows that the left interface is RT stable, 
as the density increases in the direction of acceleration.  
However, the right interface is likely unstable due to the adverse density arrangement 
(e.g., Krolik 1977, 1979; Mathews \& Blumenthal 1977; Jacquet \& Krumholz 2011; Jiang et al. 2013).  

\begin{figure}
\centering
\includegraphics[width=0.5\textwidth]{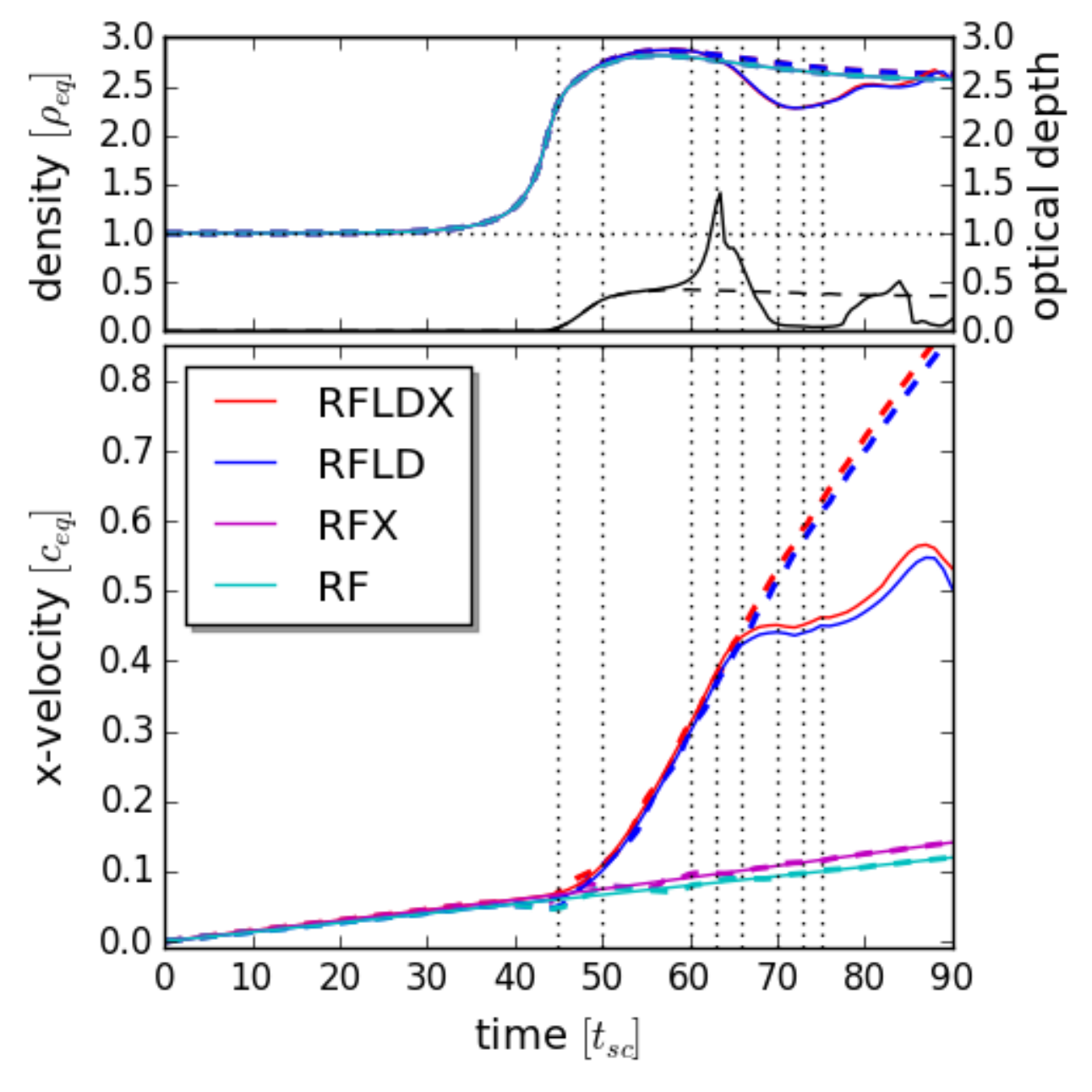}
\caption{Comparison of all runs in 1-D and 2-D.  Dashed (solid) lines denote 1-D (2-D) runs.  
In both panels, all curves nearly overlap during the nonlinear cloud formation process, which completes 
at time $\approx 50\,t_{sc}$.  In the top panel, we also verify that the cloud in run RFLDX is 
optically thin to UV radiation by calculating an estimate to the optical depth using equation 
(\ref{eq:tau_max}).  This estimate in 1-D (2-D) is given by the dashed (solid) black line.  
The dotted vertical lines mark the times corresponding to the snapshots in Fig.~4.}
\label{fig:panels}
\end{figure}

\begin{figure*}
\centering
\includegraphics[width=0.8\textwidth]{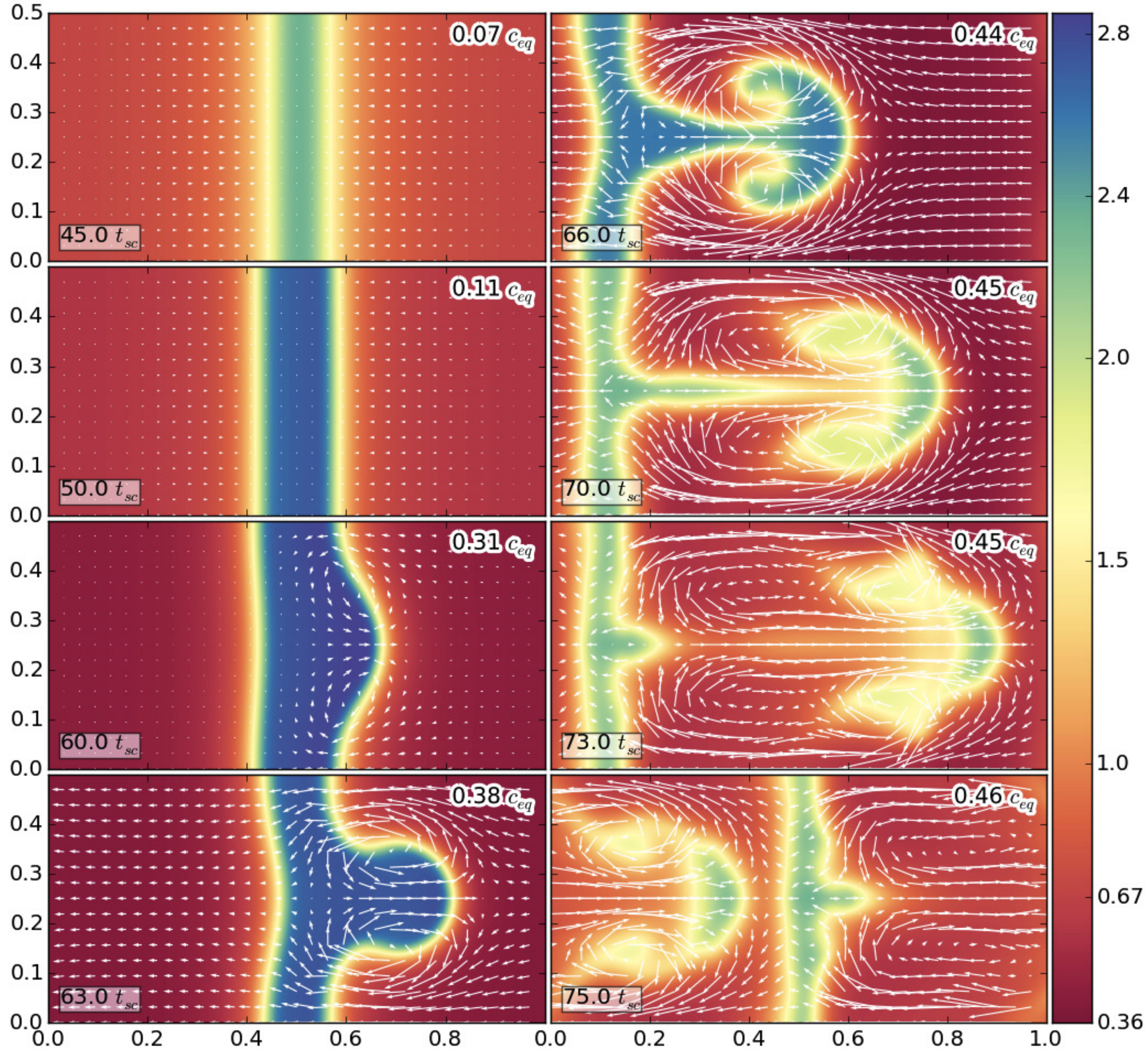}
\caption{Density snapshots of run RFLDX in 2-D in units of $\rho_{eq}$.  
The domain size is $[\lambda_x,\lambda_x/2]$ with resolution $[N_x,N_y]=[1024,512]$.  
Since the cloud continually advects through the domain boundaries, the images are manually 
aligned for visual comparison.
Velocity arrows are overlaid after subtracting the mean x-velocity of the cloud (displayed 
in the upper right corner) from $v_x$, the cloud
being defined as gas with $\rho > 1.57 \rho_{eq}$.  Time is shown in the lower left corner 
of each panel. }
\label{fig:panels}
\end{figure*}

\begin{figure}
\centering
\includegraphics[width=0.5\textwidth]{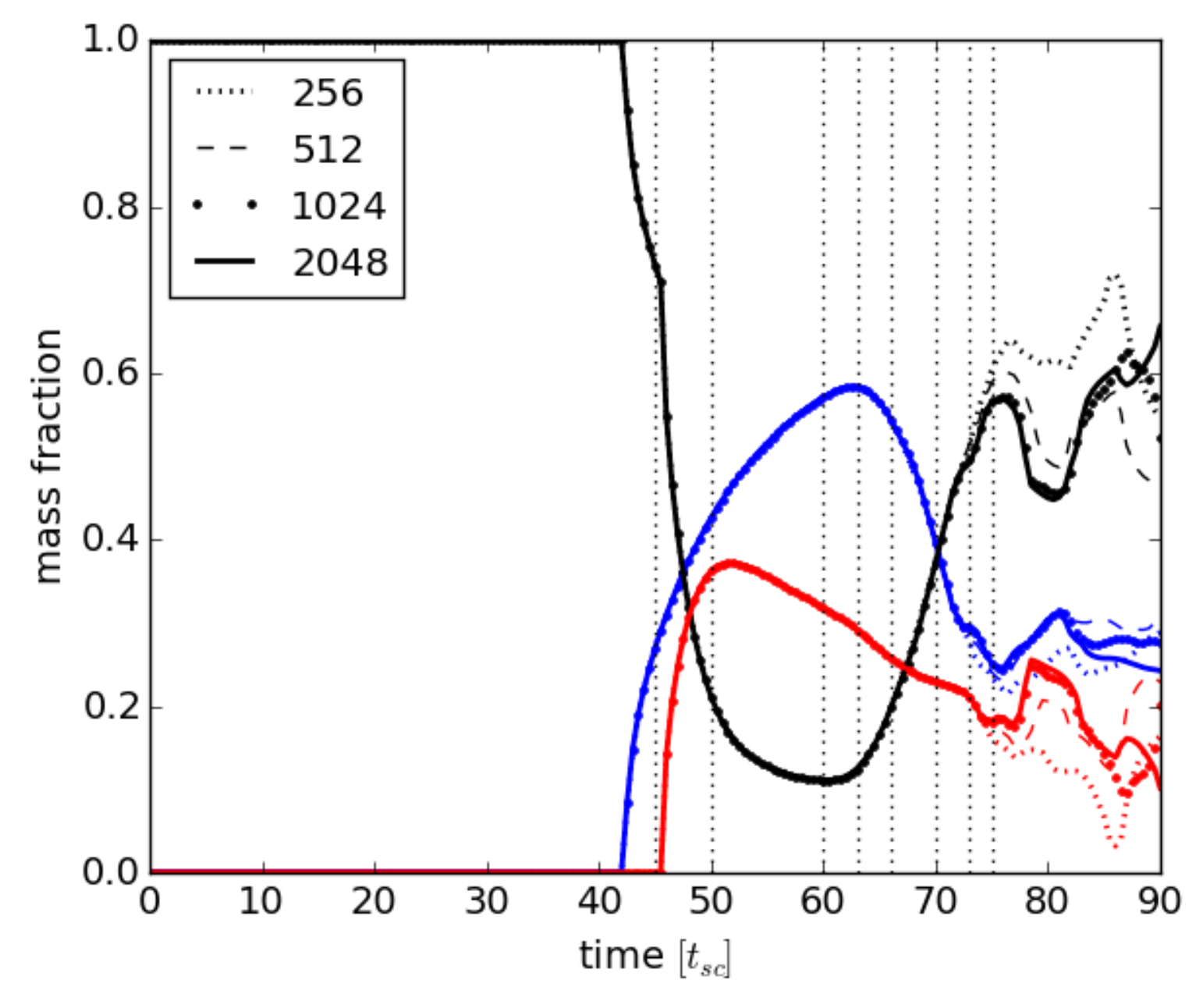}
\caption{Mass fractions for run RFLDX in 2-D at four different resolutions.  Blue, black, and 
red curves track the fraction of the gas contained in the cloud, the interfaces, and the medium, 
respectively; our method of differentiating these regions is described in \S{4.4}.  The dotted 
vertical lines mark the times corresponding to the snapshots in Fig.~4.  The mass fraction of 
the cloud monotonically increases until the RT spike becomes fully nonlinear, after which 
it begins to monotonically decrease until the spike becomes a detached structure.  At this point 
our results become resolution dependent.}
\label{fig:panels}
\end{figure}

\begin{figure}
\centering
\includegraphics[width=0.45\textwidth]{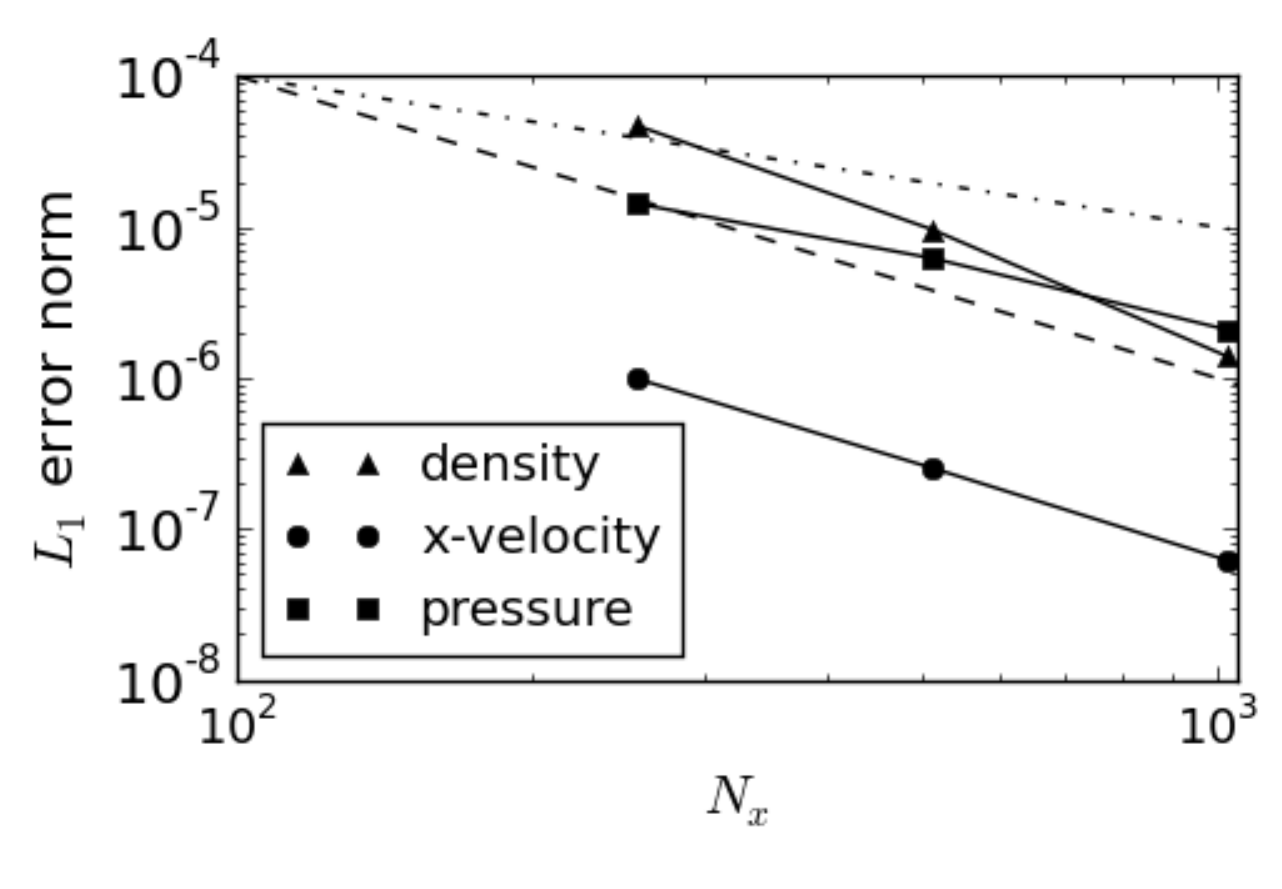}
\caption{Convergence study of run RFLDXa in 2-D, showing $L_1$ errors for resolutions 256, 512, and 1,024 
in the x-direction, with respect to the $N_x=2,048$ reference solution at time $45\,t_{sc}$.  
The dashed-dotted and dashed lines show slopes corresponding to 1st and 2nd order convergence, respectively.  
Pressure only has a convergence rate of 1, which is likely due to the reduced temporal accuracy of 
the STS scheme used to integrate the conduction term.} 
\label{fig:panels}
\end{figure}

\begin{figure}
\centering
\includegraphics[width=0.42\textwidth]{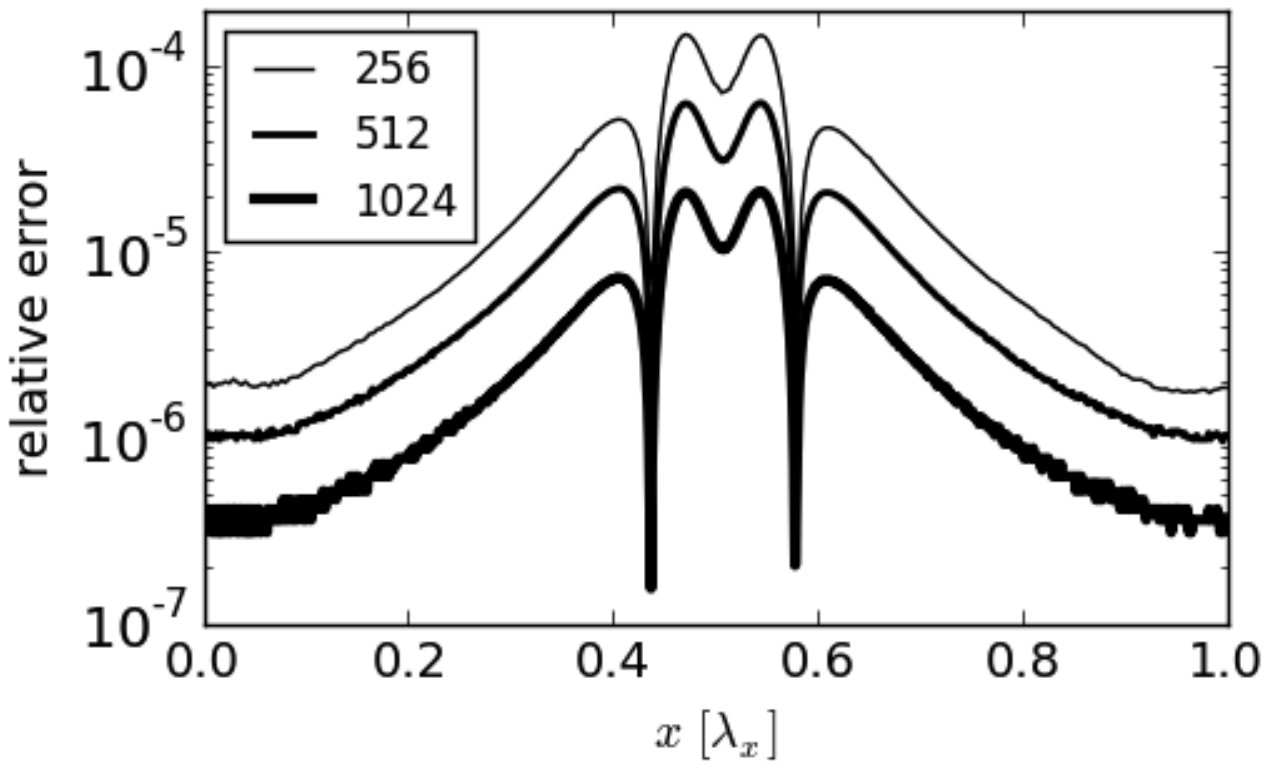}
\caption{Relative errors in pressure for run RFLDXa in 2-D at time $45\,t_{sc}$.  The `humps' centered 
around $x=0.5\,\lambda_x$ mark the conductive interfaces of the cloud, which are the most difficult regions to resolve and thus have the largest relative errors.} 
\label{fig:panels}
\end{figure}

\subsection{Results of 2-D Simulations}
The simplest extension of our 1-D simulations to 2-D is to form a cloud with a planar 
slab configuration.  The initial conditions and overall setup is as before. 
However, to fully explore 2-D effects, we now break the uniformity in 
the y-direction by introducing a perturbation with a wavelength the size of the domain in 
the y-direction (i.e. $\lambda_y = \lambda_x/2$)
and $\delta \rho = 5\times10^{-7} \rho_{eq}$.  

In Fig.~3, we present a comparison of our four 1-D runs and their 2-D counterparts, 
and we also verify our optically thin assumption.
The maximum density of the cloud versus time is plotted in the top panel.   
It is clear that there is no significant difference in any of the runs 
during the cloud formation process, which ends at $t \approx 50 t_{sc}$.  
As mentioned in \S{4.1}, for runs RFLD and RFLDX to be optically thin to UV photons, we require 
$\tau_{L,\rm max} < 1$.  We estimate $\tau_{L,\rm max}$ as
\footnote{Note that the `expanding' optical depth formula, 
$\tau_{L,\rm max} =  \sigma_e\,\eta_{\rm max}\, \rho\,v_{th}|dv/dx|^{-1}$ (see CAK) is not valid here 
because the Sobolev length $v_{th}/|dv/dx|$ is much greater than the density scale height.  That is, 
the scale over which the velocity changes by the thermal width of the line 
($v_{th} \sim 20~\rm km\,s^{-1}$) is much greater than the interface width ($\sim \lambda_F$), 
the scale over which the density (and hence opacity) changes appreciably.}
\beq \tau_{L,\rm max} =  \sigma_e \int_0^{\lambda_x}  \eta_{\rm max,90}(x) \rho(x) dx , \label{eq:tau_max}\seq
where $\eta_{\rm max,90}$ denotes only those values of  $\eta_{\rm max}$ 
in the range $[0.9\,\eta_{\rm max}, \eta_{\rm max}]$.  
We use this range to be able to identify gas
at a constant opacity in our numerical representation of  $\eta_{\rm max}$.
The black curves in the top panel of Fig.~3 show 
this estimate for $\tau_{L,\rm max}$ in both 1-D (dashed line) and 2-D (solid line).  
In the latter calculation, we consider a ray through the center of the cloud at $y=0.25\lambda_x$.  
Overall, the cloud can indeed be considered optically thin at all times, except possibly at its center 
during a very short period of the acceleration phase in 2-D, which as will be made clear below, 
coincides with when the cloud is significantly lengthened by the onset of disruptive processes. 

The bottom panel of Fig.~3 shows the average velocity of the cloud versus time, where we define 
the cloud as being the gas to the left of the grey region in Fig.~1 (i.e. $\xi < 121.2$, 
which corresponds to $\rho > 1.57 \rho_{eq}$).  It indicates that significant acceleration only 
takes place after the cloud formation process has ended, as line opacity is only activated 
once the cloud forms.  While 1-D runs of RFLD and RFLDX uniformly accelerate to supersonic speeds, 
the acceleration is suddenly halted in 2-D around $t\approx 66 t_{sc}$.  

In Fig.~4, we plot snapshots of run RLFDX, to illustrate the very different fate of 
a rapidly accelerated cloud in 2-D. 
The first frame shows that the slab formed at $t = 45\,t_{sc}$.
The initial perturbation in the y-direction grows into a slight over-density region in 
the center of 
the slab, which then undergoes greater acceleration than its surroundings and causes 
a small bulge to appear around $t\approx 55 t_{sc}$ (not shown).
Viewing this bulge as a perturbation along the surface of the slab, 
the basic criteria for the RT instability is satisfied: 
heavy fluid is pushing against light fluid.  
The same conclusion applies to runs RFX and RFLD, although for run RFX it will
take much longer (several hundred $t_{sc}$) for the bulge to grow due to the weak acceleration.  
Run RF, however, which has a constant acceleration due to 
Thomson opacity, evolves identically in 1-D and 2-D for all time; any density perturbation 
in the y-direction receives the same push as any other point in the flow. 

The remaining frames in Fig.~4 reveal how the breakup of the cloud ensues as 
the RT instability develops and soon becomes accompanied by the KH instability.  
First the bulge evolves to become mushroom-shaped, forming a structure resembling 
the classic RT `spike', which features prominent KH `rolls' at $t=66\,t_{sc}$ made 
possible by the increased relative velocity between the cloud and medium.  The halting of 
the acceleration happens around this time.  The connecting plume then disperses
(i.e. its gas is heated) as the spike separates further from the slab.  
The slab would likely disperse also, due to the mass lost to the spike, but instead 
it is somewhat thickened by the approach of the spike from the backside.  
This collision perspective is shown in the final panel at $t= 75\,t_{sc}$.  
 
The cloud is eventually either completely dispersed back into the medium, 
or coexists with it in a disordered manner in what could be called 
a clumpy flow once the vertical symmetry is lost.  The simulations 
presented here do not let us make any definitive statements because 
we noticed that our solutions become resolution dependent at about time 
$73\,t_{sc}$ for run RFLDX.  This loss of convergence is shown in Fig.~5, 
where for four different resolutions, we plot the mass fraction of 
the three components of the gas: (i) the cloud (blue), again defined as gas 
with $\xi < 121.2$ or $\rho > 1.57~\rho_{eq}$ 
(ii) the interface (black), defined as the portion of the gas in 
the grey region that is unstable, i.e. the tracks above the dashed line 
in Fig.~1 with $121 \leq \xi \leq 278$ or 
$0.68~\rho_{eq} \leq \rho \leq  1.57~\rho_{eq}$; 
(iii) the medium (red), defined as the (stable) gas with 
$\xi > 278$ or $\rho < 0.68~\rho_{eq}$.  
The mass fraction is defined as the mass of each component divided 
by the total mass and is given by
$ m = (N_xN_y)^{-1} \sum_{ij} \rho_{ij}/\rho_{eq}$, 
where $N_x$ and $N_y$ are the number of grid zones in the x and y directions 
and the sum ranges over all zones $(i,j)$ that satisfy one of 
the criteria (i)-(iii).  Once the mass fractions for resolutions 
$N_x = 1024$ and $N_x = 2048$ differ, we cannot claim 
to accurately follow the cloud's evolution.

The mass fractions provide a complementary description of the cloud evolution depicted in Fig.~4.  
The cloud appears fully formed by $t\approx50\,t_{sc}$, which coincides with the peak mass fraction 
of the medium, but mass keeps piling on until $t\approx63\,t_{sc}$.  Indeed, we observe
that the velocity arrows at $t=50\,t_{sc}$ in Fig.~4 point toward the cloud, indicating continued 
growth at the expense of the medium.  
The overall fraction of gas occupying interface regions is a minimum 
at $t\approx60\,t_{sc}$, and this is despite the overall increase in the size of interface region 
(due to the bulge) because the interfaces are narrower.   
The cloud mass fraction reaches a maximum at $t\approx63\,t_{sc}$ when the RT spike has become 
fully nonlinear, and thereafter monotonically decreases, with the mass being taken up entirely 
in the interfaces, until the RT spike becomes detached from the slab.  During this time, 
the medium continues losing mass to the interfaces.  The loss of convergence is likely due 
to the appearance of small scale structures as the cloud is disrupted.   

\subsection{2-D Convergence Study}
\label{convergence}
The resolution dependent late-time dynamics encountered above warrants demonstrating that 
our results are converged at earlier times, especially since an acceleration scheme such 
as STS was found to be necessary in 2-D due to the strict time constraint imposed by 
the CFL condition ($\Delta t \propto \Delta x^2$ due to conduction) combined with 
the long duration of the runs.  
To show that our 2-D results using STS are converged, in the top panel of Fig.~6 
we plot L$_1$ errors for run RFLDXa at $t=45\,t_{sc}$.  
This run is a modified version of run RFLDX
where we suppressed the contribution of the force due to electron scattering by a factor of 11 
(by setting the term $(1+f_{UV})$ in equation (\ref{eq:sigma_tot}) equal to 1)
to minimize the effects of advection errors not associated with the dominant term $f_{UV} M_{max}$.  

For a uniform mesh the L$_1$ error norm 
is the quantity $(N_x N_y)^{-1} \sum_{ij} R_{ij}$, where $R_{ij} = |q_{ij} - q_{ref}|$ is 
the local absolute error in some quantity $q$ at the location of zone $(i,j)$ with respect 
to some reference solution $q_{ref}$ evaluated at the same location, and the sum ranges 
over all zones in the domain.  As a reference solution we used the highest resolution simulation 
we could reasonably afford, namely $[N_x,N_y] = [2048, 1024]$.  The dashed line in Fig.~2 marks 
the slope corresponding to a convergence rate of 2, showing that the solutions for density and 
velocity converge to the reference solution in line with \textsc{Athena}'s 2nd order accurate 
spatial discretization.  However, pressure shows a reduced convergence rate, which may be due to 
the fact that the STS scheme is only 1st order accurate in time.  Interestingly, we observed 
that reducing the Courant number by a factor of 6 will result in pressure 
also showing 2nd order convergence, but our results presented here used 
the default Courant number 0.8.   

In Fig.~7 we plot relative errors for pressure, defined as $R_{ij}/|p_{ref}|$, for a horizontal 
slice through the domain at $j=0$.  This plot also demonstrates self-convergence and, 
not surprisingly, it shows that the largest errors are found at the conductive interfaces of the cloud.  

\section{Discussion}
As the next planned phase of our ongoing investigation of cloud acceleration initiated 
in Proga et al. (2014), this paper considered the cloud formation and acceleration 
processes simultaneously for the first time.
In particular, we have extended the basic theory of the nonlinear outcome of TI 
by self-consistently solving for the dynamics of optically thin gas in the presence 
of a strong radiation field.  Our resulting simulations have made it possible to 
study in detail the evolution of the isobaric condensation mode in thermally 
unstable gas from an initial perturbation to a dense, high velocity cloud.  The cloud 
forms in a radiation pressure dominated environment, but the radiation force has 
practically no effect on the cloud formation process in either 1-D or 2-D.  
The reason is simply because there is no momentum transfer from the radiation 
to the gas unless there is also sufficient opacity, and the sources of opacity 
are not activated until the cloud is formed.  

The initial motivation for this work was simply to demonstrate that the nonlinear 
phase of TI leads to a natural mechanism to produce fast clouds via acceleration due to lines.  
In so doing we were led to the inescapable conclusion that accelerated clouds undergo 
rapid deformation and are ultimately destroyed, thereby confirming long-standing 
assertions about the inevitability of cloud destruction (e.g., Mathews 1986; Krolik 1999 
and references therein).  However, we find that optically thin clouds can survive long enough 
to be accelerated to relatively high velocities and travel a significant distance of many cloud sizes, 
in contrast to investigations exploring pre-existing optically thick clouds 
(Proga et al. 2014 and references therein.)  Consequently, the best hope for cloud-based models 
of AGN is to identify robust mechanisms for continually producing clouds (e.g., in ``outflows from inflows'' 
as illustrated in simulations presented by Kurosawa \& Proga 2009 or Mo{\'s}cibrodzka \& Proga 2013).
It is therefore important to thoroughly study how clouds form via TI and how they evolve before 
disruptive processes take hold.  

We found that a rich set of dynamics unfolds during the cloud formation process.  For example, for 
run RFLDX (radiation force due to X-rays and lines) in 1-D, there are three nonlinear phases of the TI: (i)~initial growth and saturation 
$(t \approx 40 - 46 \, t_{sc})$; (ii)~evolution toward a uniformly accelerating solution 
$(t \approx 46 - 55 \, t_{sc})$ (iii)~evolution toward a thermal equilibrium state 
$(t \approx 46 - 90\, t_{sc})$.  
Figure 4 shows that phases (i) and (ii) have both completed before the RT instability can develop, 
so these phases also take place in 2-D, while phase (iii) has a different outcome in 2-D 
and the uniform acceleration cannot be maintained.   These phases will be described 
in more detail in Waters \& Proga (2015, in preparation).

Our follow up paper will also further investigate the growth of the RT and KH instabilities.  
This is an important matter since the appearance of these instabilities governs the end phase 
of TI and therefore dictates the ultimate fate of the cloud.  These instabilities develop 
after the TI saturates. 
Therefore, it should be possible to confirm the theoretical linear growth rates by conducting 
a careful numerical perturbation analysis (e.g., by introducing sinusoidal perturbations along 
the interfaces of a 2-D planar slab initialized using a 1-D solution).  That said, we did not find 
it possible to make a meaningful comparison of the growth rate of the bulge and the classical RT rate 
in the present setup.  Recalling Fig.~3, the bulge is already borderline nonlinear by $t = 60\,t_{sc}$, 
implying that the slab is still evolving thermally [i.e. in phase (iii) of nonlinear TI evolution]
during the linear RT growth regime.  This dynamical complication combined with the simplifying 
assumptions inherent in the linear theory for RT (such as constant acceleration everywhere) 
warrant using a more controlled approach for isolating the development of the individual instabilities.     

The numerical setup presented here can be used for several different exploratory studies of cloud 
formation and acceleration.  We chose the initial perturbation amplitude $\delta \rho$ in 
the y-direction to be $10^{-2}$ that in the x-direction in order to arrive at a slab configuration.  
With an equal ratio (and equal growth rates), a round cloud will be formed 
in 2-D instead.  Multiple clouds can be formed 
using higher wavenumber perturbations.  Cloud fragmentation can be studied by decreasing 
the ratio $t_{th}/t_{sc}$.  Classical evaporation (Cowie \& McKee 1977; Balbus 1985) can be 
explored by arranging for the Field length to exceed the size of the cloud. This large range of 
initial configurations would be difficult or impossible to construct otherwise, which illustrates 
an obvious advantage of studying cloud acceleration via the formation process, namely that the internal 
gas dynamics is self-consistently treated.  Indeed, we find that pressure equilibrium with 
the surrounding medium is naturally maintained, cloud interfaces form with a width determined 
by the conductivity, the radiation and drag forces reach a balance so that hydrostatic equilibrium 
is established in the reference frame of the cloud, and the cloud reaches a thermal equilibrium state 
in which heating by conduction is balanced by line cooling, in agreement with Begelman \& McKee (1990).   
Moreover, the equilibrium location on the S-curve largely determines the cloud density, temperature, 
and opacity before it is accelerated, while plotting the evolutionary tracks on the $T-\xi$ plane 
has shown itself to be a useful tool both for understanding the time evolution of the cloud and 
for characterizing the components of 
the gas.\footnote{Simulations demonstrating this can viewed online at www.physics.unlv.edu/astro/pw15sims.html}

Taking into consideration the numerical requirements involved in this study, multi-dimensional simulations 
will likely be constrained to only explore values of $\xi_3 \lesssim 300$ for the S-curve used here, 
thereby limiting the overall density contrast of clouds to $\chi \approx 18$.  This limitation arises 
due to the need to resolve interfaces, as simulations of clouds formed via TI will not be converged 
unless the conductive interfaces are resolved (Koyama \& Inutsuka 2004).  Previous numerical studies 
using pre-existing clouds without thermal conduction and with unresolved interfaces were able 
to explore much higher density contrasts such as $\chi = 50$ (McCourt et al. 2014) and 
even $\chi \sim 10^4$ (Krause et al. 2012).  We found that a realistic $\kappa \propto T^{5/2}$ 
would require upwards of $N_x = 4,096$ zones (i.e. at least three levels of refinement if adaptive 
mesh refinement is employed) even for our modest density contrast of $\chi \approx 8$, as would 
values of $\chi \gtrsim 30$ with continued use of a constant conductivity.  This rapid steepening 
of the interfaces with either increased $\chi$ or realistic $\kappa(T)$ results in ever smaller 
transition regions between the cloud and the medium but does not imply that the role of thermal 
conductivity becomes less important.  As a consequence, the conduction time step would be so 
small at these resolutions that even STS schemes would become impractically slow, necessitating 
the use of implicit techniques.   Such extensions to this work may be needed to assess, for example, if the 
the timescale for cloud destruction is sensitive to the slope of the interfaces, i.e. if it decreases with steeper 
density gradients, as was found to be the case when a cloud is disrupted 
by the passage of a shock (Nakamura et al. 2006).   

Future work is also needed to address important aspects of cloud formation and acceleration 
neglected here.  Magnetic fields can play a dominant role in both, as they can prevent the gas from 
condensating in the first place (e.g., Mathews \& Doane 1990), they may help accelerate clouds 
through confinement (e.g., Arav \& Li 1994), and they can significantly effect cloud dynamics 
due to the effects of anisotropic conduction (Choi \& Stone 2011).  
Other environmental factors such as the presence of Coriolis forces, gravity, 
winds, shocks, etc. could be incorporated to build a more global model of 
cloud dynamics for comparison with observations.  However, it is crucial
to model 3-D effects, as both compression and expansion of clouds can be 
enhanced by a large factor, and this can affect the timescale for 
cloud destruction. A direct extension of this study designed to form initially 
spherical clouds in 3-D would violate our optically thin assumption and 
therefore requires a full RHD treatment using methods that can accurately 
capture shadows (e.g., Davis et al. 2012; Jiang et al. 2012).  Such methods, 
which have already been employed by Proga et al. (2014), would also allow 
investigation of the so-called line-deshadowing instability 
(Owocki \& Rybicki 1984) that could lead to 
changes in the structure and time-dependence of the clouds.  

\section{Acknowledgments}
This work was supported by NASA under ATP grant NNX11AI96G.  
Our simulations were performed on the Eureka cluster at UNLV's 
National Supercomputing Center for Energy and the Environment.  
We thank J. Stone for developing the \textsc{Athena} code and making it public,
as well as for discussions about this work.  We are grateful to the referee 
for constructive comments that helped us improve this work. TW thanks 
Chad Meyer for discussions regarding the use and accuracy of STS schemes. 

\newpage
\appendix
The net cooling function ${\mathcal L}$ that defines our S-curve is comprised of 
four heating and cooling rates. 
The analytic expressions for $L_{\rm ff}$ and $G_{C}$ are
\beq L_{\rm ff} = \f{2^5\pi e^6}{3h m_e c^3} \sqrt{ \f{2\pi k_B T}{3m_e} } Z^2 \bar{g}_B= 3.3\times10^{-27}\sqrt{T} \,\,{\rm [erg\: cm^3 s^{-1} ]}, \text{   and} \seq
\beq G_C = \f{k_B \sigma_e}{4\pi m_e c^2}\xi\, T_X \left(1- 4\f{T}{T_X} \right) = 8.9 \times 10^{-36} \xi\,T_X\left(1 - 4\f{T}{T_X}\right)  \:{\rm [erg\,\, cm^3 s^{-1} ]}, \seq
where $m_e$ and $e$ are the mass and charge of an electron, $h$ is Planck's constant, $Z$ is the ion atomic number, $\bar{g}_B$ is an averaged Gaunt factor, and $T_X = 10\,\rm{keV}/k_B$.
The analytical fits from Blondin (1994) in our notation read
\begin{align}
L_{bb} &=  \delta \left[ 1.7 \times 10^{-18} \exp\left(- \frac{1.3 \times 10^5}{T}\right) \xi^{-1} T^{-1/2} + 10^{-24} \right]   {\, \, \rm [erg\,\, cm^3 s^{-1} ]}, \label{eq:Lbb}  \text{   and}   \\
G_X &= 1.5 \times 10^{-21} \xi^{1/4} T^{-1/2} \left(1-\f{T}{T_X}\right)  {\rm [erg\,\, cm^3 s^{-1} ]}.
\end{align}
Here, $\delta$ is a parameter introduced by Blondin (1994); setting $\delta < 1$ mimics 
reducing the strength of line cooling when relaxing his assumption of optically thin gas.  
We keep $\delta = 1$ since we assume optically thin clouds.  In \S{\ref{equations}}, we refer to 
the heating part of $G_X$, which is $G_{X,h} = 1.5 \times 10^{-21} \xi^{1/4} T^{-1/2} ~{\rm [erg\,\, cm^3 s^{-1} ]} $.
Finally, it is important to note that Blondin's photoionization calculations were revisited 
and independently verified using XSTAR by Dorodnitsyn et al. (2008) for an incident AGN power 
law spectrum.  Their analytical fits differ from the above only by a minor modification 
to equation (\ref{eq:Lbb}), which Dorodnitsyn et al. (2008) report had no significant 
dynamical effects on their simulation results.  

For line-driving, we use equation (17) from Proga (2007) 
\beq
\log{k_{\rm{CAK}}} = \left\{ \begin{array}{ll}
                   -0.383  & {\rm for} ~\,~     \log T \leq 4 \\
-0.630 \log{T}+2.138  & {\rm for} ~\,~ 4 < \log T \leq4.75 \\
-3.870 \log{T}+17.528 & {\rm for} ~\,~     \log T > 4.75 \\
\end{array}
\right.
\seq
and equation (19) from Stevens \& Kallman (1990):
\begin{equation}
\log \eta_{\rm max} = \left\{ \begin{array}{ll}
6.9~\exp(0.16\,\xi^{0.4})
& {\rm for}~\,~
\log \xi \leq 0.5 \\
 & \\
9.1 \exp(-7.96\times10^{-3}\xi)
& {\rm for} ~\,~
\log\xi > 0.5 . \\
\end{array}
\right.
\label{eq:SK90eq19}
\end{equation}

\end{document}